\documentclass[a4paper]{jpconf}

\include{jpconf11} 
 \pdfoutput=1 

\usepackage{graphicx}

\newcommand{\eh}[1]{\,\mathrm{#1}}

\newcommand{\degree}{$^{\circ}$}

\begin{document}

\title{Cosmic Ray Physics with the IceCube Observatory}

\author{H.~Kolanoski (for the IceCube Collaboration)}

\address{Humboldt-Universit\"at  zu Berlin and DESY}

\ead{Hermann.Kolanoski@desy.de}

\begin{abstract}
The IceCube Neutrino Observatory  with its 1-km$^3$ in-ice detector and the 1-km$^2$ surface detector (IceTop) constitutes a three-dimensional cosmic ray detector well suited for general cosmic ray physics.
Various measurements of cosmic ray properties, such as energy spectra, mass composition and anisotropies, have been obtained from analyses of air showers at the surface and/or  atmospheric muons in the ice. 
\end{abstract}

\section{Introduction}
The IceCube Neutrino Observatory  \cite{achterberg06,Kolanoski_HLT_icrc2011} is a detector situated in the ice of the geographic South Pole at a depth of about 2000 m. The observatory is primarily designed to measure neutrinos from below, using the Earth as a filter to discriminate against muon background induced by cosmic rays (neutrino results are reported elsewhere in these proceedings \cite{kappes_HLT_ecrs2012}). 
IceCube also includes an air shower array on the surface called IceTop extendind IceCube's capabilities for cosmic ray physics.
Construction of IceCube Neutrino Observatory was completed in December  2010.  

IceCube can be regarded as a cubic-kilometer scale three-dimensional cosmic ray detector with the air showers (mainly the electromagnetic component) measured by the surface detector IceTop and the high energy muons and neutrinos measured in the ice. In particular the measurement of the electromagnetic component in IceTop in coincidence with the high energy muon bundle, originating from the first interactions in the atmosphere, has a strong sensitivity to composition. Here IceCube offers the unique possibility to clarify the cosmic ray composition and spectrum in the range between about 300 TeV and 1 EeV, including the `knee' region and a possible transition from galactic to extra-galactic cosmic rays. 

\section{Detector}

\begin{figure}
	\centering	\includegraphics[width=0.62\textwidth]{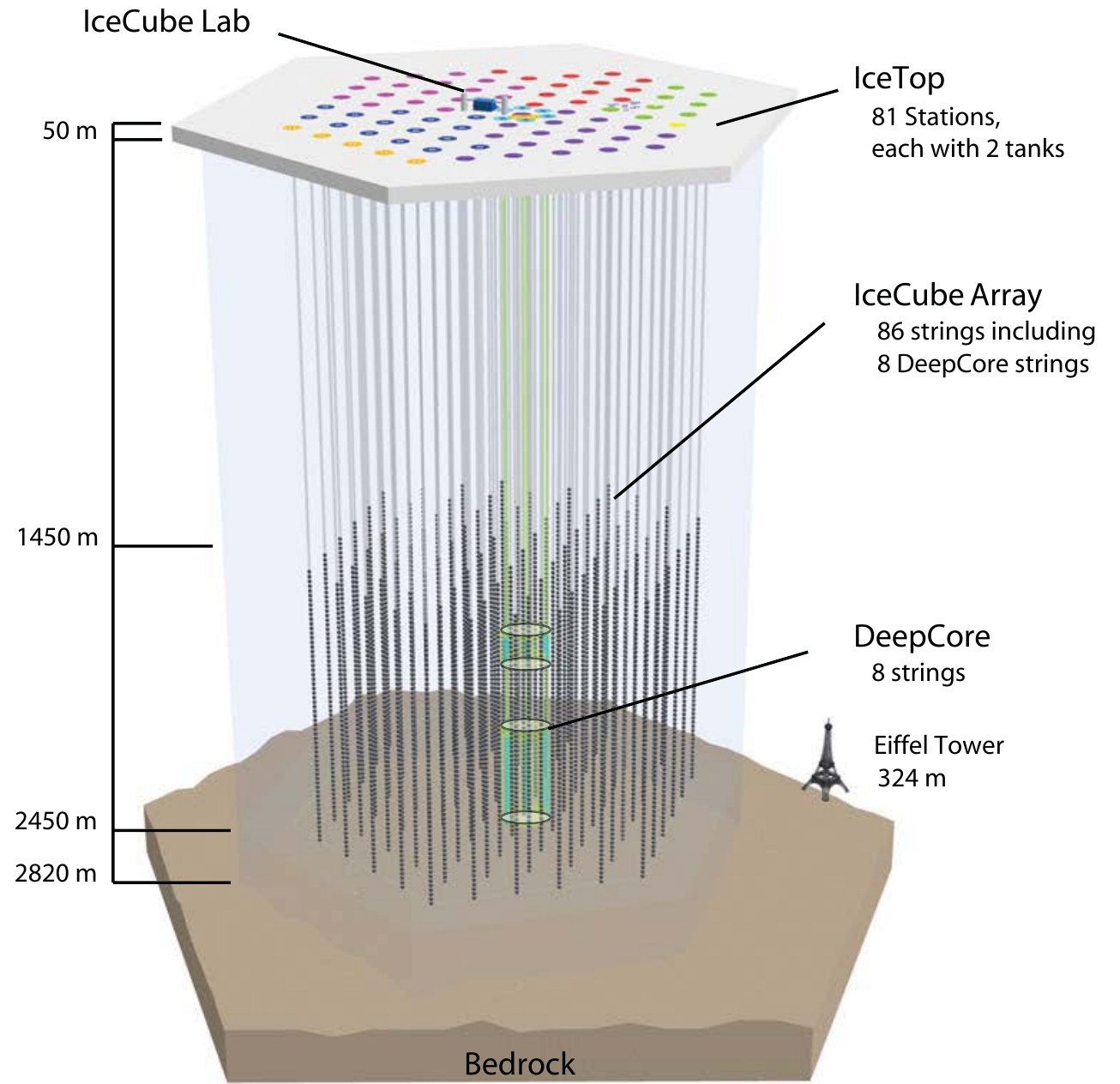}\hfill
\includegraphics[width=0.37\textwidth]{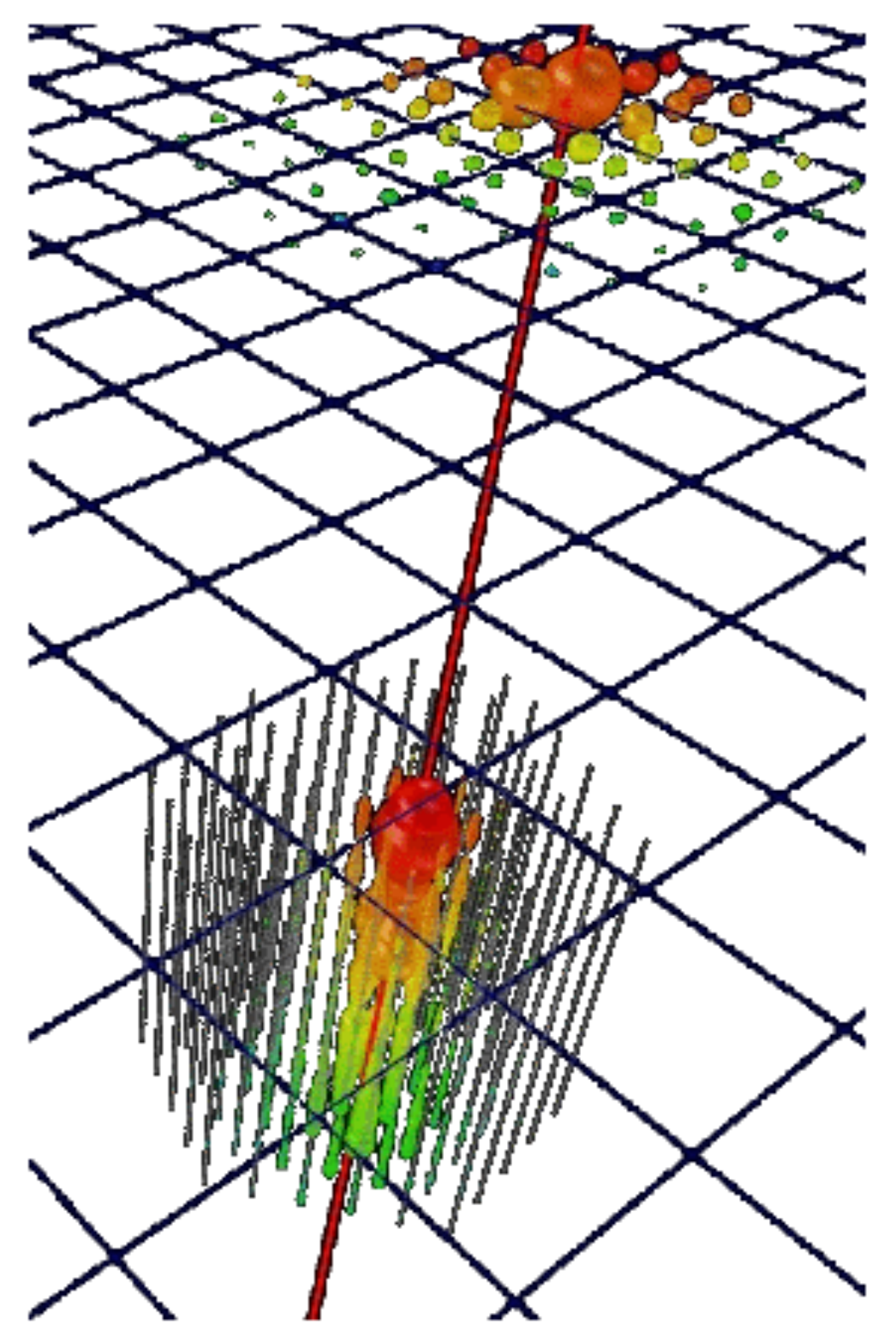}

	\caption{Left: The IceCube detector with its components DeepCore and IceTop in the final configuration (January 2011). In this paper we present data taken with the still incomplete detector. We will refer to the configuration as IC79/IT73, for example, meaning 79 strings in IceCube and 73 stations in IceTop. The final detector has the configuration IC86/IT81. Right: View of a cosmic ray event which hits IceTop and IceCube.   The size of the colored spots is proportional to the signal in the DOMs, the colors encode the signal times, separately for IceCube and IceTop. }\label{fig:I3Array}

\end{figure}

\paragraph{IceCube:}
The main component of the IceCube Observatory is an array of 86 strings equipped with 5160 light detectors in a volume of 1 km$^3$ at a depth between 1450\,m and 2450\,m (Fig.\,\ref{fig:I3Array}). The nominal IceCube string spacing is 125 m on a hexagonal grid. A part of the detector, called DeepCore, is more densely instrumented resulting in a lower energy threshold.

Each string, except those of DeepCore, is equipped with 60 light detectors, called `Digital Optical Modules' (DOMs), each containing a $10''$ photo multiplier tube (PMT) to record the Cherenkov light of charged particles traversing the ice. In addition, a DOM houses complex electronic circuitry supplying signal digitisation, readout, triggering, calibration, data transfer and various control functions. The most important feature of the DOM electronics is the recording of the analog waveforms in $3.3\eh{ns}$ wide bins for a duration of $422\eh{ns}$.
 With a coarser binning a `fast ADC' extends the time range to 6.4\,$\mu$s. 

\paragraph{IceTop:}
The 1-km$^2$ IceTop air shower array \cite{ITDet-IceCube:2012nn} is located above IceCube at a height of 2835\,m above sea level, corresponding to an
atmospheric depth of about 680 g/cm$^2$. It consists of 162 ice Cherenkov tanks, placed at 81 stations mostly near the IceCube strings (Fig.\,\ref{fig:I3Array}). 
In the center of the array, a denser station distribution forms an in-fill array with a lower energy threshold (about 100\,TeV). Each station comprises two cylindrical tanks, 10 m apart, 
 with an inner diameter of $1.82\eh{m}$ and filled with ice to a height of $90\eh{cm}$. 

Each tank is equipped with two DOMs which are operated at different PMT gains to cover linearly a dynamic range of about $10^5$ with a sensitivity to a single photoelectron (the thresholds, however, are around 20 photoelectrons). DOMs, electronics and readout scheme are the same as for the in-ice detector.

\section{Cosmic Ray spectrum}\label{sec:spectrum}
\begin{figure}
\begin{minipage}[b]{0.48\textwidth}	
	\centering		\includegraphics[width=1.0\textwidth]{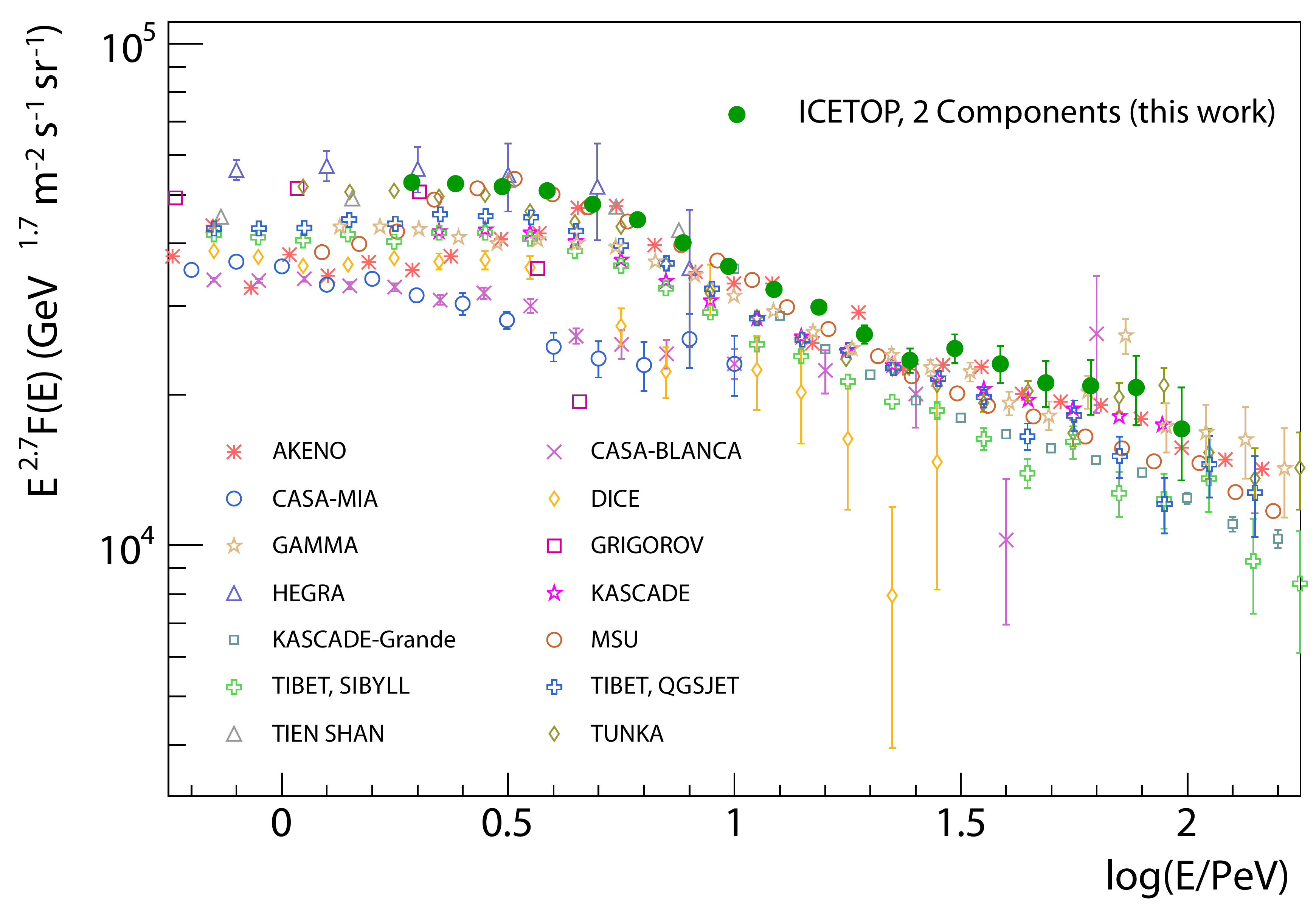}\vspace{-1mm}
\caption{Energy spectrum measured with IceTop  in the 26-station configuration assuming a two-component model for the primary mass composition. The IceTop data are compared to other measurements in this energy range (references in \cite{IT26-spectrum_Abbasi:2012wn}).\\{\ }}\label{fig:IT26_spectrum-v2_2}
	\end{minipage}\hfill
	\begin{minipage}[b]{0.48\textwidth}
	\centering
	\includegraphics[width=1.00\textwidth]{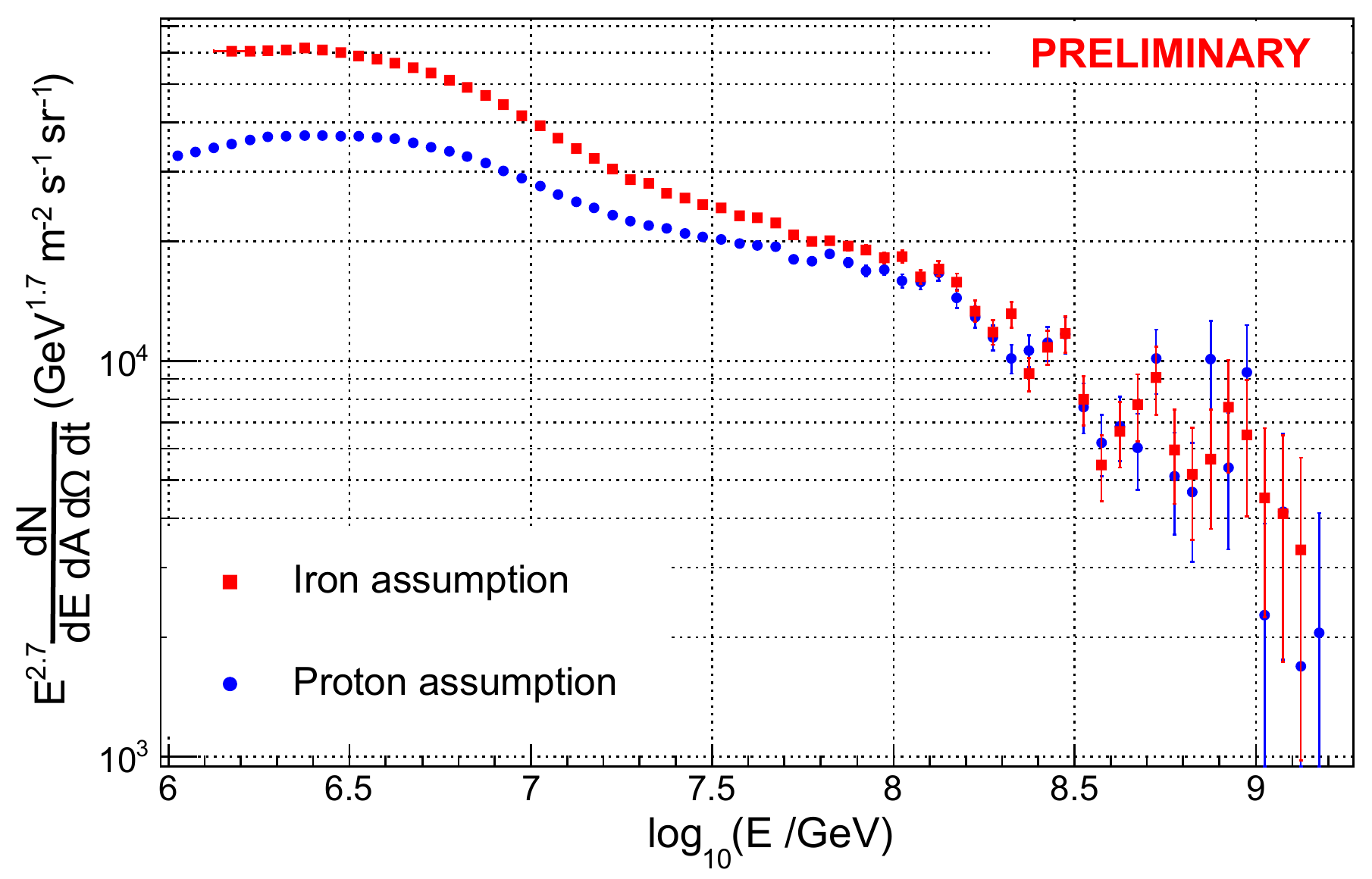}
	\caption{First evalution of one year of data taken with the 73-station configuration of IceTop in 2010. The events were required to have more than 5 stations and zenith angles in the range $\cos\theta \geq 0.8$. The spectrum is shown for the two assumptions `pure proton' and `pure iron' for the primary composition. }	\label{fig:IT73-spectrum-p-Fe}
\end{minipage}
\end{figure}
\begin{figure}
	\centering
\includegraphics[width=0.43\textwidth]{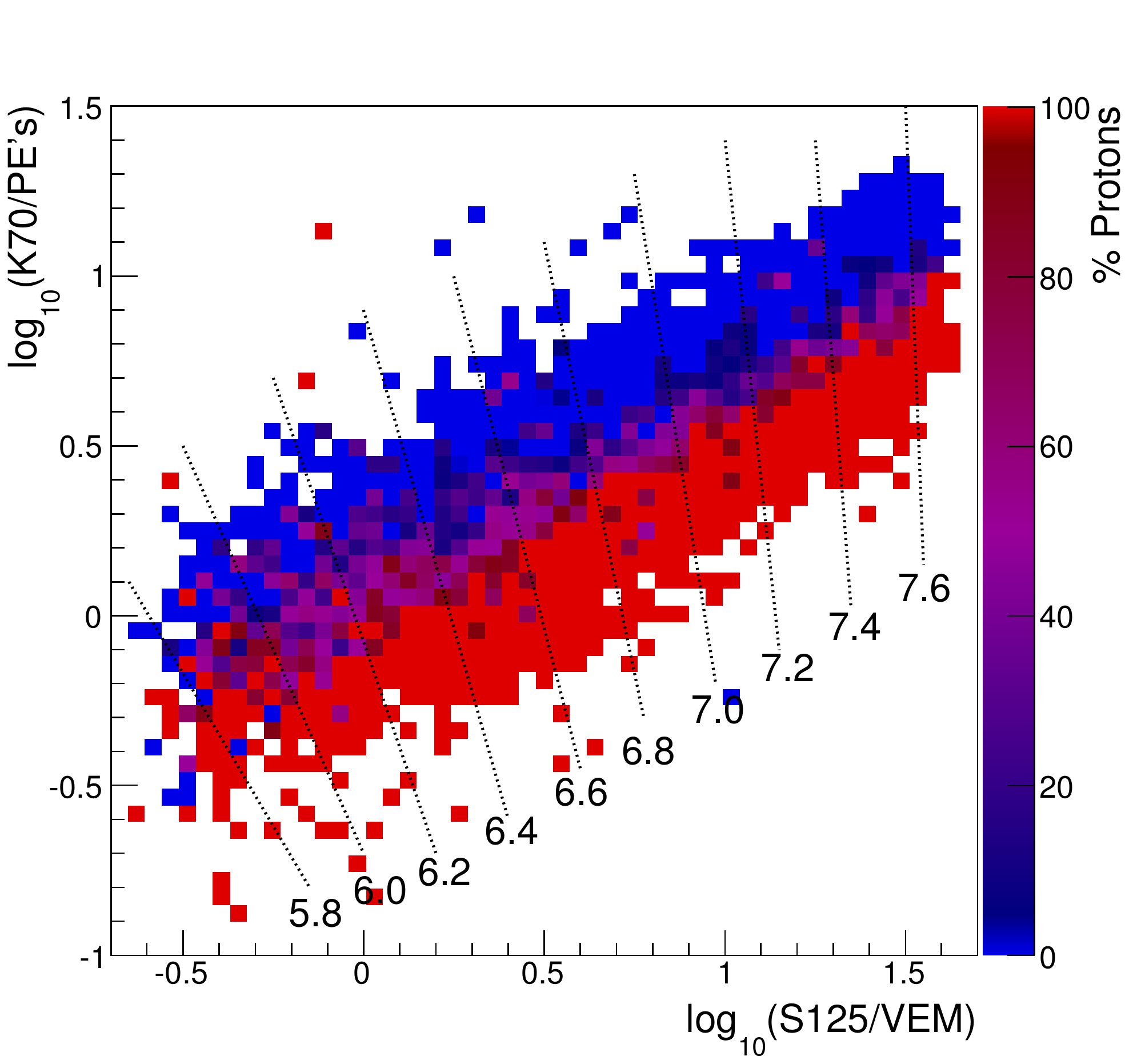}	\hfill
\includegraphics[width=0.54\textwidth]{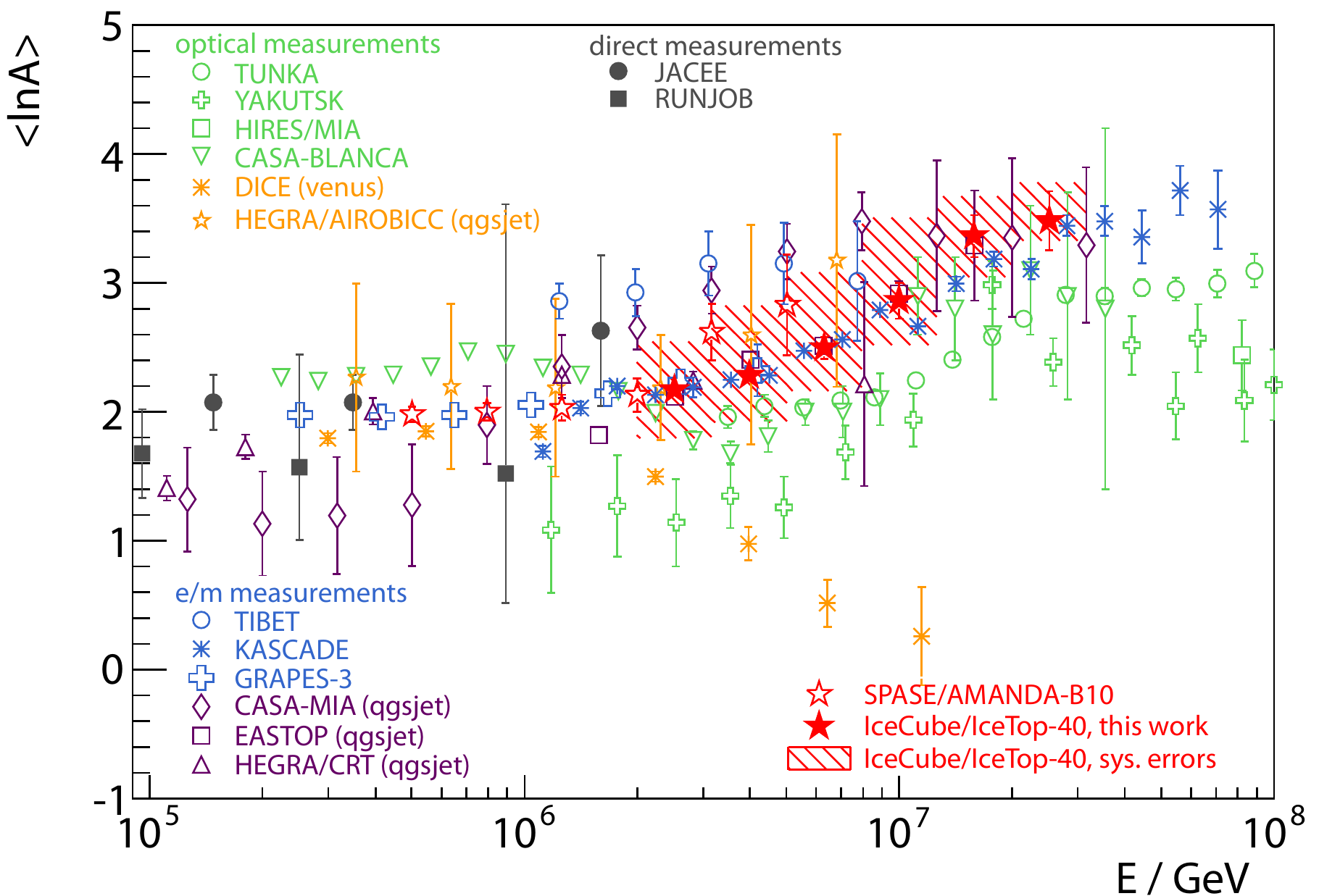}\vspace{-4mm}
	\caption{Composition analysis (IC40/IT40 configuration) \cite{ITIC40-composition_Abbasi:2012}. Left: Simulated correlation between the energy loss of the muon bundles in the ice (K70) and the shower size at the surface (S125) for proton and iron showers. The shading indicates the percentage of protons over the sum of protons and iron in a bin. The lines of constant primary energy are labeled with the logarithms of the energies. Right: IceCube result for the average logarithmic mass of primary cosmic rays compared to other measurements (references in \cite{ITIC40-composition_Abbasi:2012}). }
	\label{fig:composition_ITIC40}
\end{figure}

Figure \ref{fig:IT26_spectrum-v2_2} shows the energy spectrum from 1 to 100 PeV \cite{IT26-spectrum_Abbasi:2012wn} determined from 4 month of data taken in the IT26 configuration in 2007. 
The relation between the measured shower size and the primary energy is mass dependent.  Good agreement of the spectra in three zenith angle ranges was found for the assumption of pure proton and a simple two-component model (see \cite{IT26-spectrum_Abbasi:2012wn}). For zenith angles below 30\degree , where the mass dependence is smallest, the knee in the cosmic ray energy spectrum was observed at about 4.3\,PeV with the largest uncertainty coming from the composition dependence (+0.38\,PeV and -1.1\,PeV).
The spectral index changes from 2.76 below the knee to 3.11 above the knee. There is an indication of a flattening of the spectrum above about 20\,PeV which was also seen by the experiments GAMMA \cite{Gamma-Garyaka:2008gs}, Tunka \cite{Kuzmichev_HLT_ecrs2012} and Kaskade-Grande \cite{Haungs_HLT_ecrs2012}.

A first preliminary evaluation of IceTop data from the 2010/11 season with 79 IceCube strings and 73 IceTop stations is shown in Fig.~\ref{fig:IT73-spectrum-p-Fe}.    

\section{Cosmic ray composition}\label{sec:composition}

As mentioned in the introduction, the combination of the in-ice detector with the surface detector offers a unique possibility to determine the spectrum and mass composition of cosmic rays from about 300 TeV to 1 EeV. 
 
 The first such analysis exploiting the IceTop-IceCube correlation was done on a small data set corresponding to only one month of data taken with about a quarter of the final detector  for energies from 1 to 30 PeV \cite{ITIC40-composition_Abbasi:2012}. From the measured input variables,  shower size and  muon energy loss  (Fig.\,\ref{fig:composition_ITIC40}, left), the primary energy and mass was determined using a neural network.  The resulting average logarithmic mass is shown in Fig.\,\ref{fig:composition_ITIC40}, right. These results are still dominated by systematic uncertainties, such as the energy scale of the muons in IceCube and the effects of snow accumulation on the IceTop tanks. 

 A similar analysis of  IceTop-IceCube coincidences is in progress using the IC79/IT73 data set taken in 2010 (the energy spectrum obtained with these data is displayed in Fig.~\ref{fig:IT73-spectrum-p-Fe}). The studies indicate that there will be enough statistics for composition analysis up to about 1 EeV. 
 
The systematic uncertainties related to the models can be reduced by including different mass sensitive variables, like zenith angle dependence of shower size \cite{IT26-spectrum_Abbasi:2012wn}, muon rates in the surface detector and shower shape variables (see discussion in \cite{Kolanoski_HLT_icrc2011}).

\section{PeV-gamma rays}\label{sec:pevgamma}

\begin{figure}
	\centering
		\includegraphics[width=0.48\textwidth]{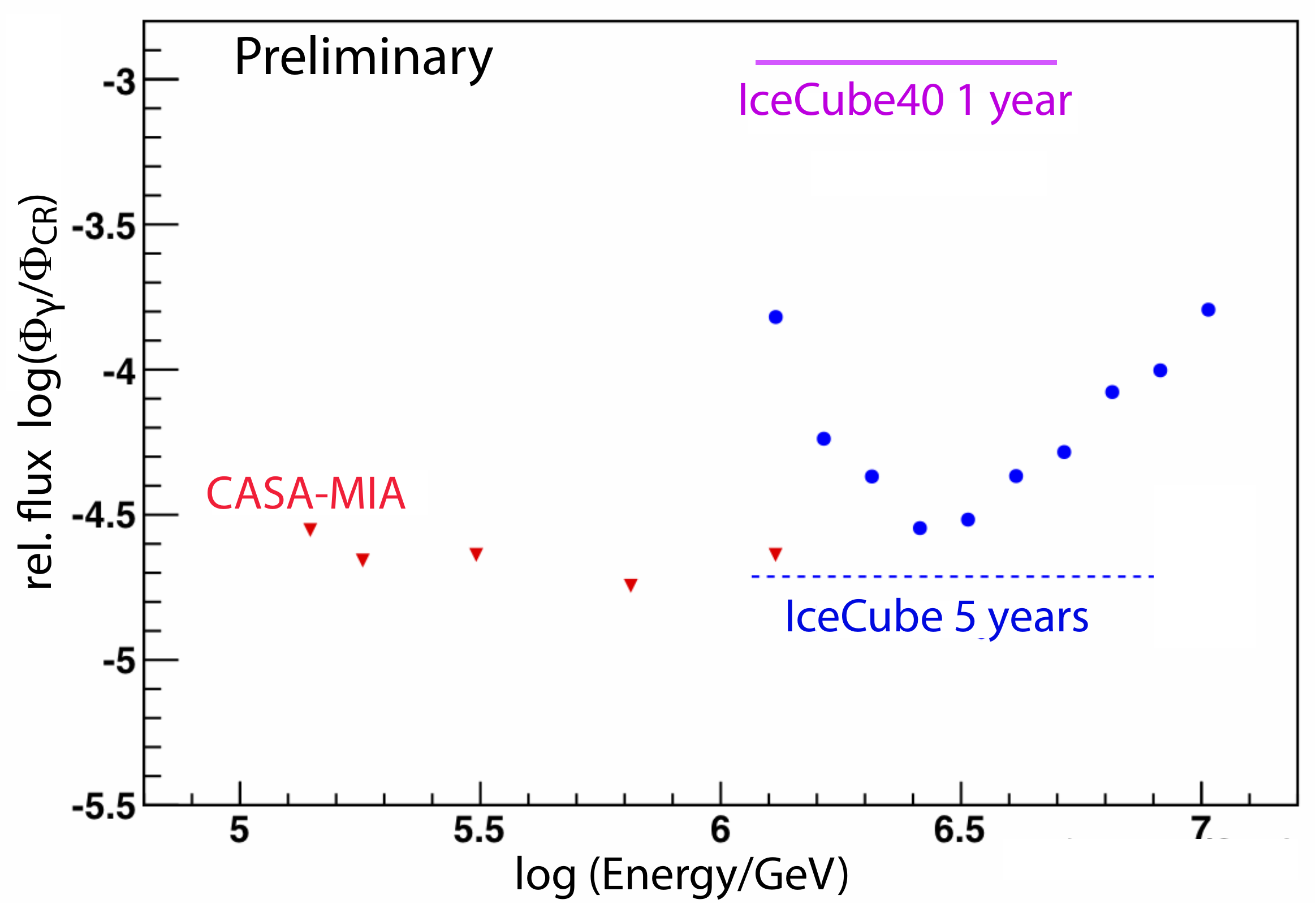}\hfill
\begin{minipage}[b]{0.48\textwidth}	\caption{Limits on the diffuse gamma ray flux relative to the cosmic ray flux from a region within 10\degree\ from the Galactic Plane (IC40/IT409, purple line). The plot includes also the only other available limits from CASA-MIA \cite{CASA-MIA-Chantell:1997gs} and the expected one-year sensitivity for the complete IceCube detector (blue dashed line for the whole covered energy range, blue dots for smaller energy bins).\\{\ }}\label{fig:pevgammarays_limits}
\end{minipage}
\end{figure}

IceCube can efficiently distinguish PeV gamma rays from the background of cosmic rays by exploiting coincident in-ice signals as veto. Gamma-ray air showers have a much lower muon content than cosmic ray air showers of the same energy. Candidate events are selected from those showers that lack a signal from a muon bundle in the deep ice.
 Results of one year of data, taken in the IC40/IT40 configuration are shown in Fig.~\ref{fig:pevgammarays_limits} \cite{Stijn_icrc2011}. The projected
gamma-ray sensitivity of the final detector is also given. 

\section{Transient events}\label{sec:transients}

Transient events such as sun flares or gamma ray bursts, if they generate very high fluxes of low energy particles, could be observed as general rate increases above the noise level in the IceTop DOMs even if they could not be detected individually. This was first demonstrated with the observation of the Dec 13, 2006 Sun flare event \cite{Sun-flare-Abbasi08}. The detector readout has since then been setup such that counting rates could be obtained at different thresholds allowing to unfold cosmic ray spectra during a flare \cite{Takao_IT_icrc2011}.

\section{Atmospheric muons in the ice}\label{sec:muons_inice}
In this section analyses of atmospheric muons in IceCube (without requiring air shower detection in IceTop) are presented. The related atmospheric neutrinos, an irreducible background for cosmic neutrino search, are discussed elsewhere in these proceedings \cite{kappes_HLT_ecrs2012}.

\subsection{Muon spectrum and composition}

\begin{figure}
	\centering
		\includegraphics[width=0.48\textwidth]{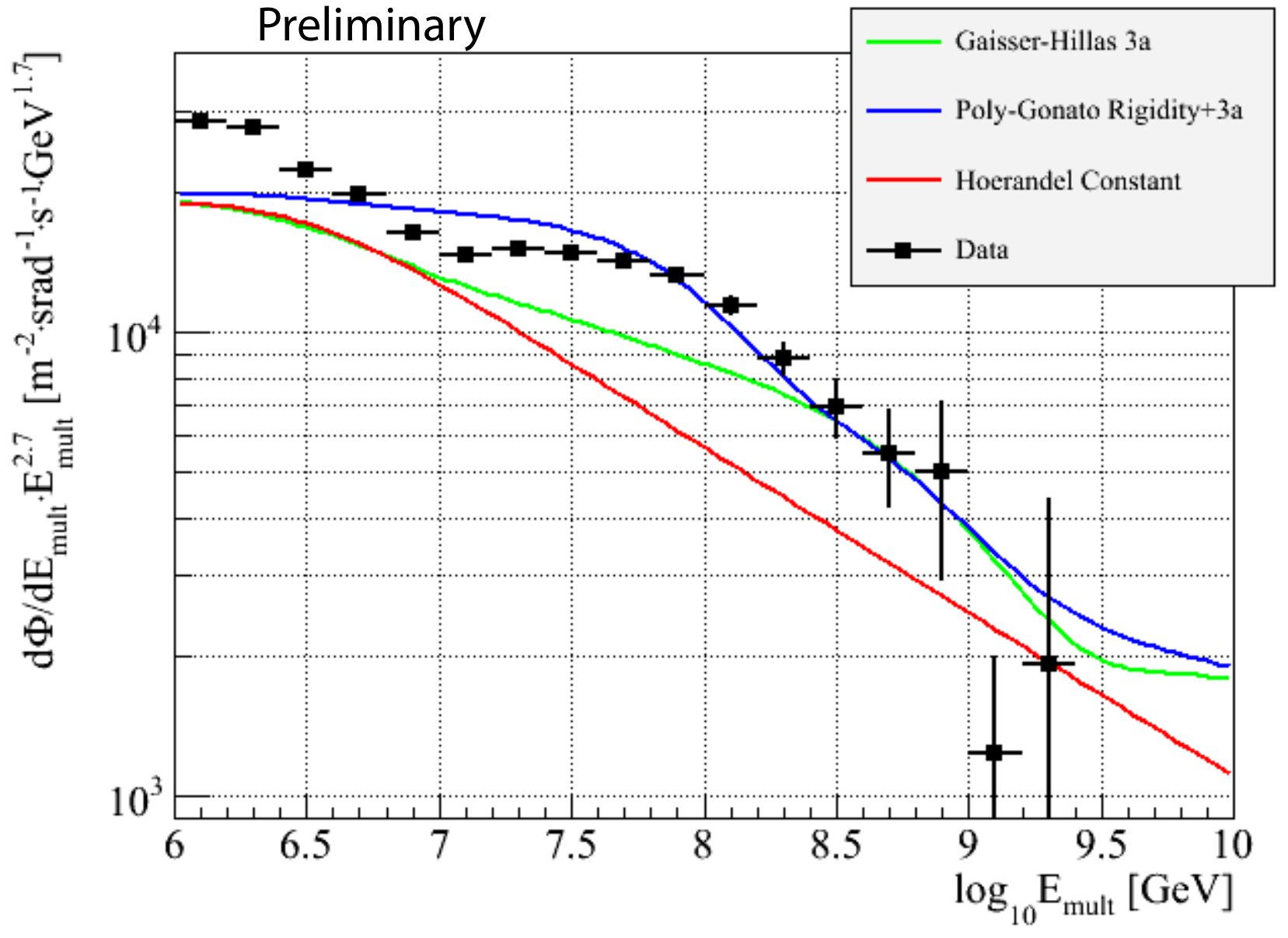}
		\hfill
	\begin{minipage}[b]{0.48\textwidth}
	\caption{Energy spectrum of primary cosmic rays obtained from muon bundles in IceCube. The energy estimator $E_{mult}$ is derived from the measured muon multiplicity in the bundles which is composition dependent. See explanation in the text. \\ {\ } }\label{fig:Muon-bundle-spectrum}\end{minipage}
\end{figure}

Atmospheric muon and neutrino spectra measured with IceCube probe shower development of cosmic rays with primary energies above about 10 TeV. To penetrate to the IceCube depth and be detectable the muons have to have energies above about 500 GeV. Methods have been developed to distinguish single, high-energetic muons by their stochastic energy loss \cite{Berghaus_icrc2011} from muon bundles with rather smooth energy deposition.  Figure \ref{fig:Muon-bundle-spectrum} shows a cosmic ray spectrum derived from an analysis of muon bundles. The flux is plotted against an energy estimator, $E_{mult}$, which is derived from the measured muon multiplicity in the bundles using the empirical formula $N_{\mu} \sim A^{0.23} E^{0.77}$ with iron as reference nucleus ($A=56$). The data are compared to the predictions from different models. None of the models matches particularly well, especially not at low energies (where threshold effects might cause some experimental uncertainty). The data indicate that some additional component at higher energies is required, for example the extra-galactic `mixed component' in the model ``Gaiser-Hillas 3a''  (see Fig.\,1 in \cite{Gaisser:2012zz} and discussion in \cite{Berghaus_isvhecri_2012}). There is also an interesting flattening observable above about 10 PeV which might be connected to the flattening observed  in the same region in the IceTop spectrum (Figs.\ \ref{fig:IT26_spectrum-v2_2} and \ref{fig:IT73-spectrum-p-Fe}) and by other experiments (see Section \ref{sec:spectrum}).

This analysis is complementary to the composition analysis 
and can be exploited to test the consistency of models in a wide energy range from well below the knee to above some EeV.

\subsection{Muons with high transverse momenta}    
 At high energies the muons reach the in-ice detector in bundles which are, for primaries above about 1\,PeV, collimated within radii of the order of some 10\,m. Most of the muons stem from the soft peripheral collisions with little transverse momentum transfer. Perturbative QCD calculations, however, predict the occurrence of muons with higher transverse momenta in some fraction of the events.  
 
\begin{figure}
\begin{minipage}[b]{0.48\textwidth}\includegraphics[width=1.0\textwidth]{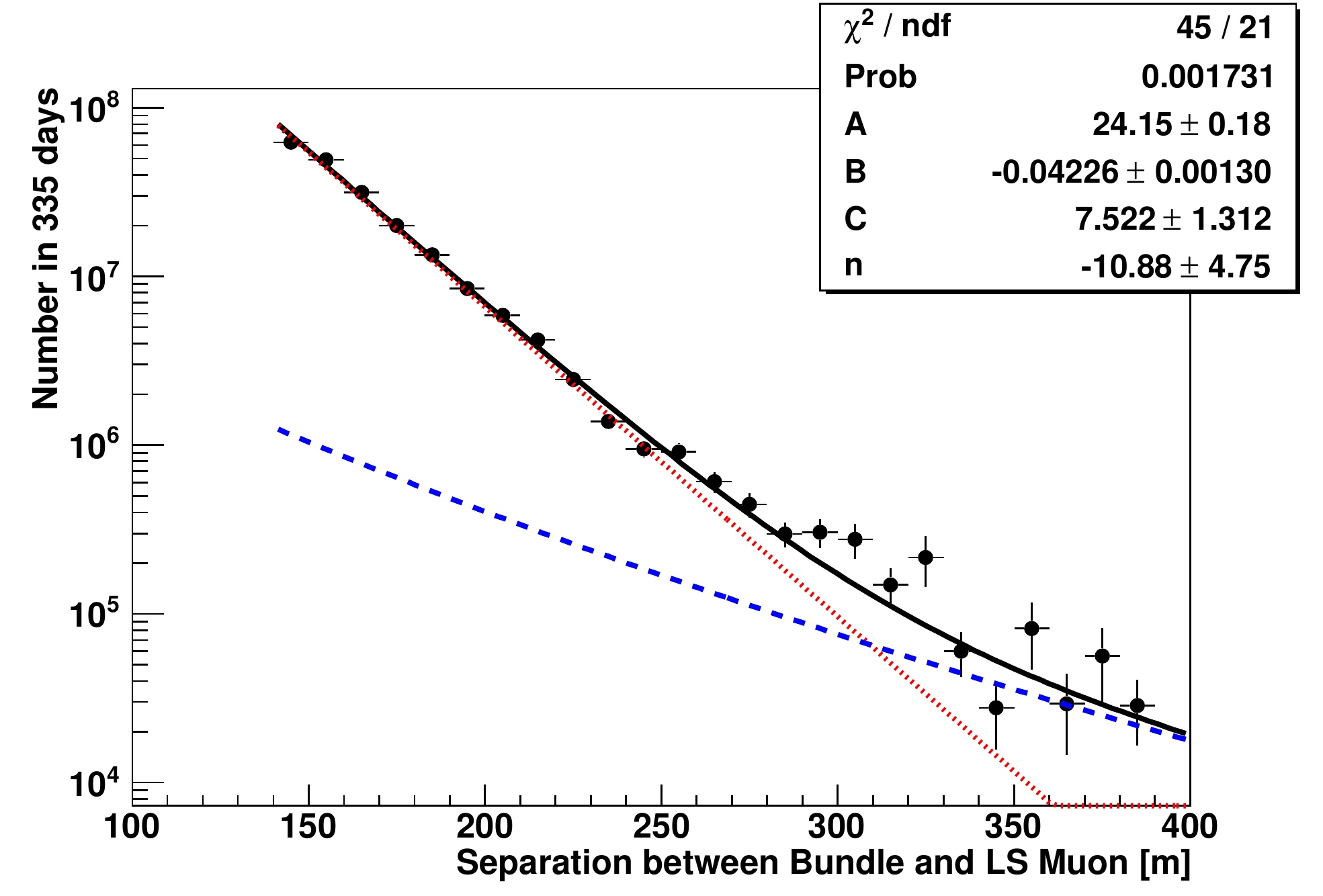}
\caption{\label{fig:LS-dist} The lateral separation distribution at sea level, fitted by a function containing an exponential part (dotted red line) and a power law (blue line).}\end{minipage}\hfill
\begin{minipage}[b]{0.48\textwidth}\includegraphics[width=1.0\textwidth]{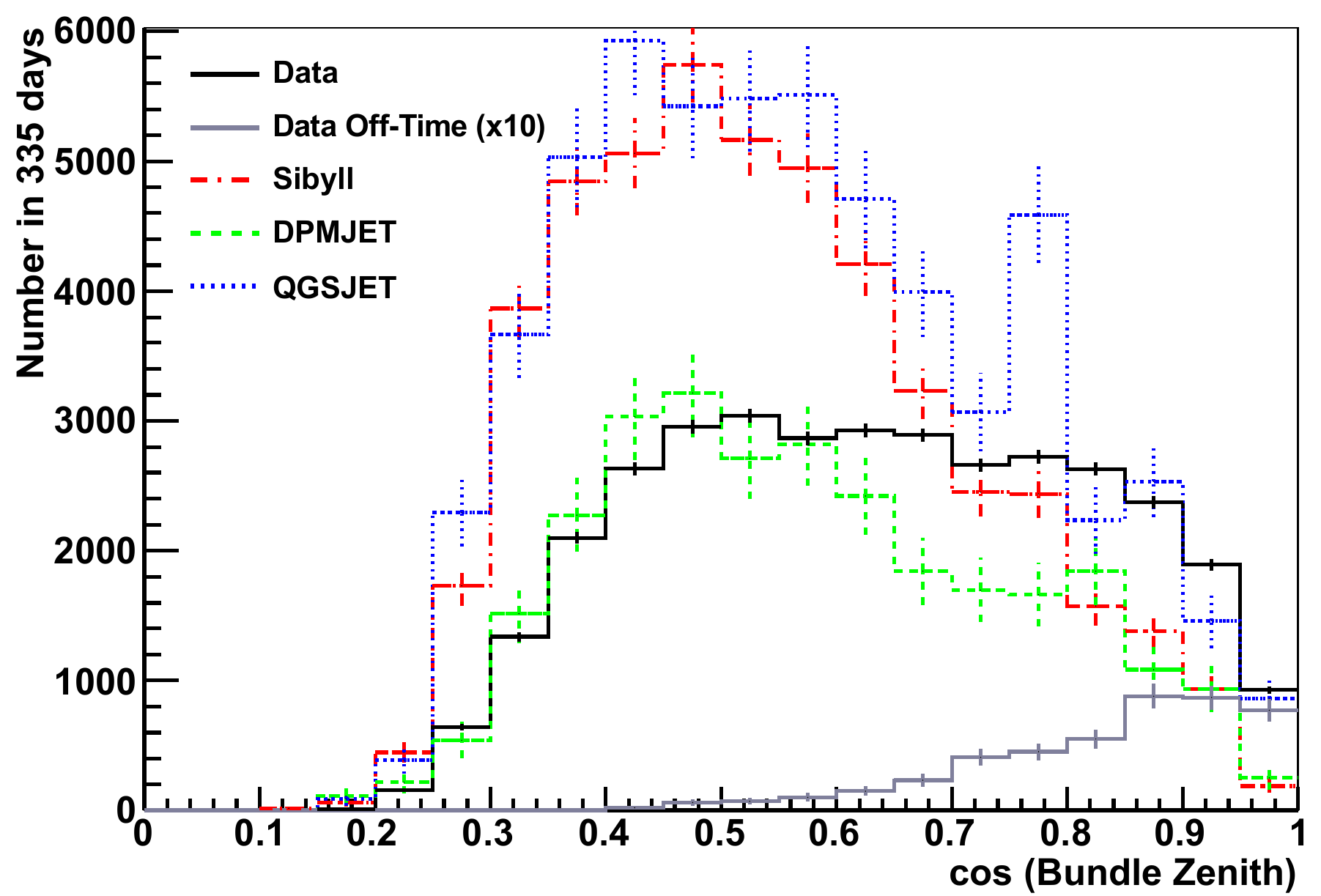}
\caption{\label{fig:LS-zenith}  The cosine distribution of the directions of  bundles with laterally separated muons compared to simulations using commonly used interaction models.}\end{minipage}
\end{figure}

Large transverse momenta of muons show up in a lateral separation from the muon bundle. In Fig.~\ref{fig:LS-dist} this lateral distribution obtained from IC59 data \cite{LSMuons-Abbasi:2012he} is shown along with a fit by an exponential plus a power law function. The power law part indicates the onset of hard scattering in this regime of $p_T \approx 2-15$ GeV/c, as expected from perturbative QCD. However, the zenith angular dependence shown in Fig.~\ref{fig:LS-zenith} cannot be described by the commonly used models Sibyll and QGSJET while it is reasonably reproduced by DPMJET. The reasons for these differences have to be understood and could have  important implications for air shower simulations.

\section{Cosmic ray anisotropy}\label{sec:anisotropy}

IceCube collects large amounts of cosmic ray muon events, about $10^{11}$ events in every year of running with the full detector. 
These events have been used to study cosmic ray anisotropies on multiple angular scales, for the first time in the Southern sky \cite{Abbasi_anisotropy:2010mf,Abbasi_anisotropy:2011ai,Abbasi_anisotropy:2011zka}. 
\begin{figure}
	\centering
\includegraphics[width=1.0\textwidth]{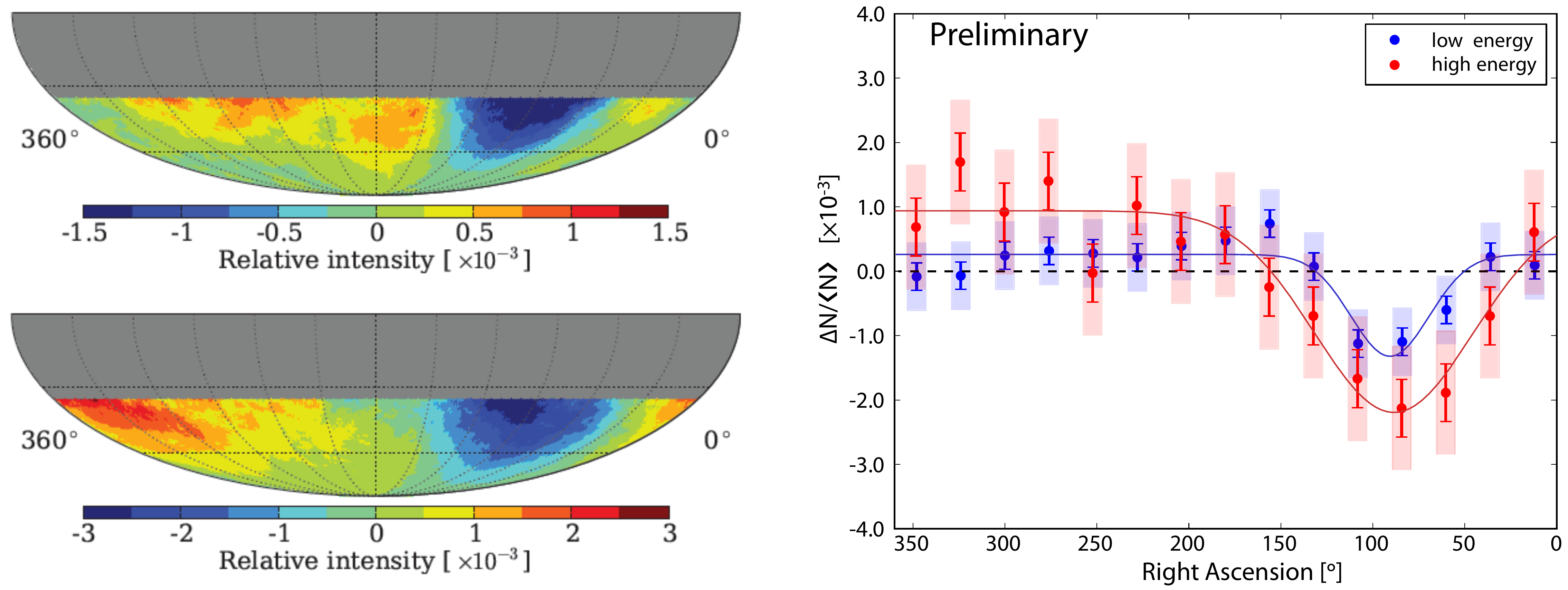}
	\caption{Left: Relative intensity maps for the low-energy (top) and high-energy (bottom) data sets. Right: Projections of the maps unto right ascension in the declination band -75\degree\ to -25\degree . In the projection plot, the error bars are statistical while the colored boxes indicate the systematic uncertainty. The curves are empirical fits.}
	\label{fig:IT-aniso-RelInt}
\end{figure}
While the previous analyses exploited data from the in-ice muons only, now also first results from data taken with IceTop are available. The advantage of using IceTop is a better energy resolution which allows a finer energy binning if statistics is sufficient.

Figure \ref{fig:IT-aniso-RelInt} shows skymaps of relative intensities determined from IceTop data for  primary energies centered around 400 TeV and 2 PeV (still with a rather coarse binning). The data were taken over 3 years in the configurations IT59, IT73, IT81. The 400 TeV data confirm the in-ice observations \cite{Abbasi_anisotropy:2011zka}, in particular the change in phase compared to the 20-TeV observations. The new result is that at 2\,PeV the anisotropy as a function of right ascension has a similar shape as at 400\,TeV but becomes apparently stronger.
 
As yet the anisotropies observed on multiple angular scales  and at different energies have not found an explanation. Theoretical explanations like local magnet fields affecting the cosmic ray streams and/or nearby sources of cosmic rays are discussed. The determination of the energy dependence of anisotropies will be crucial for validating such explanations.

\section{Conclusion}

The presented results on cosmic ray properties, such as energy spectra, mass composition and anisotropies, demonstrate the high, partly unique, potential of the IceCube Observatory for studying cosmic rays physics. The IceCube/IceTop system covers an energy range from well below the knee to the expected onset of an extra-galactic component.  

\section*{References}
\bibliography{ecrs_pcr2_hlt_Kolanoski}

\providecommand{\newblock}{}
\begin{thebibliography}{10}
\expandafter\ifx\csname url\endcsname\relax
  \def\url#1{{\tt #1}}\fi
\expandafter\ifx\csname urlprefix\endcsname\relax\def\urlprefix{URL }\fi
\providecommand{\eprint}[2][]{\url{#2}}

\bibitem{achterberg06}
Achterberg A {\em et~al.\/} ({IceCube Collaboration}) 2006 {\em Astropart.
  Phys.\/} {\bf 26} 155--173

\bibitem{Kolanoski_HLT_icrc2011}
Kolanoski H ({for IceCube Collab.}) {\em Proc.\ of ICRC 2011, Beijing\/}
  (\textit{Preprint} \eprint{astro-ph.HE/1111.5188})

\bibitem{kappes_HLT_ecrs2012}
Kappes A ({for the IceCube Collaboration}) {\em Proc.\ of ECRS 2012, Moscow\/}

\bibitem{ITDet-IceCube:2012nn}
Abbasi R {\em et~al.\/} (IceCube Collaboration) 2012  Submitted to NIM A
  (\textit{Preprint} \eprint{1207.6326})

\bibitem{IT26-spectrum_Abbasi:2012wn}
Abbasi R {\em et~al.\/} ({IceCube Collaboration}) 2012  Submitted to ApP
  (\textit{Preprint} \eprint{1202.3039})

\bibitem{ITIC40-composition_Abbasi:2012}
Abbasi R {\em et~al.\/} ({IceCube Collaboration}) 2012  Submitted to ApP
  (\textit{Preprint} \eprint{1207.3455})

\bibitem{Gamma-Garyaka:2008gs}
Garyaka A {\em et~al.\/} 2008 {\em J.Phys.G\/} {\bf G35} 115201
  (\textit{Preprint} \eprint{0808.1421})

\bibitem{Kuzmichev_HLT_ecrs2012}
Kuzmichev L ({for the Tunka Collaboration}) {\em Proc.\ of ECRS 2012, Moscow\/}

\bibitem{Haungs_HLT_ecrs2012}
Haungs A ({for the KASCADE-Grande Collaboration}) {\em Proc.\ of ECRS 2012,
  Moscow\/}

\bibitem{Stijn_icrc2011}
Buitink S ({for IceCube Collab.}) {\em Proc.\ of ICRC 2011, Beijing\/}
  (\textit{Preprint} \eprint{astro-ph.HE/1111.2735})

\bibitem{CASA-MIA-Chantell:1997gs}
Chantell M {\em et~al.\/} (CASA-MIA Collaboration) 1997 {\em Phys.Rev.Lett.\/}
  {\bf 79} 1805--1808

\bibitem{Sun-flare-Abbasi08}
Abbasi R {\em et~al.\/} ({IceCube Collaboration}) 2008 {\em Astrophys.\ J.\
  Lett.\/} {\bf 689} 65

\bibitem{Takao_IT_icrc2011}
Kuwabara T and Evenson P ({for IceCube Collab.}) {\em ICRC 2011, Beijing\/}
  (\textit{Preprint} \eprint{astro-ph.HE/1111.2735})

\bibitem{Berghaus_icrc2011}
Berghaus P ({for IceCube Collab.}) {\em Proc.\ of ICRC 2011, Beijing\/}
  (\textit{Preprint} \eprint{astro-ph.HE/1111.2735})

\bibitem{Gaisser:2012zz}
Gaisser T~K 2012 {\em Astropart.Phys.\/} {\bf 35} 801--806

\bibitem{Berghaus_isvhecri_2012}
Berghaus P ({for IceCube Collab.}) {\em Proc.\ of ISVHECRI 2012, Berlin\/}

\bibitem{LSMuons-Abbasi:2012he}
Abbasi R {\em et~al.\/} (IceCube Collaboration) 2012  Submitted to PRD
  (\textit{Preprint} \eprint{astro-ph.HE/1208.2979})

\bibitem{Abbasi_anisotropy:2010mf}
Abbasi R {\em et~al.\/} (IceCube Collaboration) 2010 {\em Astrophys.J.\/} {\bf
  718} L194 (\textit{Preprint} \eprint{1005.2960})

\bibitem{Abbasi_anisotropy:2011ai}
Abbasi R {\em et~al.\/} (IceCube Collaboration) 2011 {\em Astrophys.J.\/} {\bf
  740} 16 (\textit{Preprint} \eprint{1105.2326})

\bibitem{Abbasi_anisotropy:2011zka}
Abbasi R {\em et~al.\/} ({IceCube Collaboration}) 2012 {\em Astrophys.J.\/}
  {\bf 746} 33 (\textit{Preprint} \eprint{1109.1017})

\end{thebibliography}
\end{document}